# The Nu Class of Low-Degree-Truncated, Rational, Generalized Functions. Ic. IMSPE-optimal designs with circular-disk prediction domains

Nikoloz Chkonia[1] and Selden Crary[2*]

*1. Computer Science Department*
*Hamilton College, Clinton, NY, USA*

*2. Palo Alto, CA, USA*

**Abstract**

This paper is an extension of Part I of a series about Nu-class generalized functions. A method is presented for computing the integrated mean-squared prediction error ($IMSPE$) in the design of computer experiments, when the prediction domain is a circular disk. The method is extensible to more than two factors, to prediction domains other than squares or disks, and to a variety of assumed covariance functions. Three example optimal designs, under Gaussian covariance with known hyperparameters, are found using the method: an $n = 1$ design centered on the disk; an $n = 2$, continuously rotatable design with assumed inversion symmetry about the center of the disk; and an $n = 4$, twin-point design similar to the $n = 4$ twin-point design observed previously for a square prediction domain [1]. The four-point design on the disk demonstrates that non-round boundaries are not a prerequisite for the occurrence of twin-point optimal designs.

> *And the circle - they will square it some fine day (some fine day).*
> *-- Gilbert and Sullivan's operetta Princess Ida*

## 1. Introduction

The present paper is Part Ic of a Roman-numeralled series Parts I through V reporting research into Nu-class multifunctions [2]. We answer the challenge that twin-point designs, such as those reported in [3,1] and shown in Fig. 1.1, below, cannot arise on round, coincident design and prediction domains. In Sec. 4, we develop a technique to compute the $IMSPE$ objective function on circular-disk prediction domains, and in Secs. 5-7, we find three $IMSPE$-optimal designs on round, coincident design and prediction domains, including a twin-point design, thus answering the challenge.

Our research assumes the customary Cartesian covariance kernel [3], and is distinct from Padonou and Roustant's study of designs for polar Gaussian processes that uses a combination of a covariance kernel for the radius and a covariance kernel for the angle [4]. Readers interested in modeling physical processes involving, for example, diffusion from the center of a disk, or a rotation best carried out in cylindrical coordinates, are referred to [4].

*Authors' names listed alphabetically

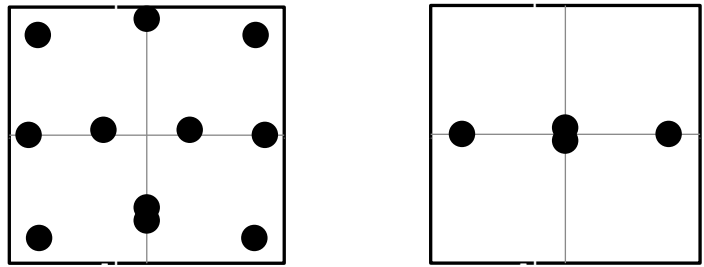

Fig. 1.1. Two $IMSPE$-optimal twin-point designs, both on the coincident, design and prediction domain $[-1,1]^2$ are shown: (left) an $n = 11$ design from [3]; and (right) an $n = 4$ design from [1]. Off-axis and non-axis-aligned twins have also been discovered [4]. In the limit of infinite optimal-design-search precision, the separations of the twins are zero, but the separations are displayed as finite, for clarity of their locations and orientations. Statistical practitioners have considerable latitude in separating a pair of twin points, while maintaining their orientation, due to the concept, "seam of ($IMSPE$) indifference," introduced in [1].

## 2. Outline

## 3. Integrals and a matrix identity

### 3.1 Elementary integral

Dwight 350.01 [5] gives almost literally:

$$\int \sqrt{a^2 - x^2}\, dx = \frac{x\sqrt{a^2 - x^2}}{2} + \frac{a^2}{2}\sin^{-1}\left(\frac{x}{a}\right), \qquad a \geq 0. \tag{3.1.1}$$

Thus, $\int_{-1}^{1} \sqrt{1 - x^2}\, dx = \left[\frac{x\sqrt{1-x^2}}{2} + \frac{1}{2}\sin^{-1}(x)\right]\Big|_{-1}^{1} = \frac{1}{2}\left(\frac{\pi}{2} + \frac{\pi}{2}\right) = \frac{\pi}{2}.$

### 3.2 Basic, unnormalized probability integral

$$I \equiv \int_{x=-W}^{x=W} e^{-\theta(a-x)^2}\, dx = \sqrt{\frac{\pi}{4\theta}}\{erf[\sqrt{\theta}\,(W + a)] + erf[\sqrt{\theta}\,(W - a)]\}. \tag{3.2.1}$$

*Demonstration:*

Changing variables:

$t = \sqrt{\theta}\,(a - x)$ gives $dt = -\sqrt{\theta}\, dx$, and the lower and upper limits of the integral transform to $t = \sqrt{\theta}\,(a + W)$ and $t = \sqrt{\theta}\,(a - W)$, respectively.

$I = -\frac{1}{\sqrt{\theta}}\left[\int_{\sqrt{\theta}\,(a+W)}^{\sqrt{\theta}\,(a-W)} e^{-t^2}\, dt\right].$ The integrand is an even function, so

$I = -\frac{1}{\sqrt{\theta}} \cdot \frac{1}{2}\left[\int_{\sqrt{\theta}\,(a+W)}^{-\sqrt{\theta}\,(a+W)} e^{-t^2}\, dt - \int_{\sqrt{\theta}\,(a-W)}^{-\sqrt{\theta}\,(a-W)} e^{-t^2}\, dt\right].$ By the definition of $erf(x)$,

$I = -\sqrt{\frac{\pi}{4b}}\{erf[-\sqrt{\theta}\,(a + W)] - erf[-\sqrt{\theta}(a - W)]\}.$ Because $erf(x)$ is an odd function,

$I = \sqrt{\frac{\pi}{4\theta}}\{erf[\sqrt{\theta}\,(W + a)] + erf[\sqrt{\theta}\,(W - a)]\}.$

### 3.3 More unnormalized probability integrals

The following two integrals are generalizations of Eqs. 3.2.1.

$$\int_{x_{2;k}-\Delta}^{x_{2;k}+\Delta} e^{-\theta(a-x)^2}\, dx = \sqrt{\frac{\pi}{4\theta}}\{erf[\sqrt{\theta}(\Delta - x_{2;k} + a)] + erf[\sqrt{\theta}(\Delta + x_{2;k} - a)]\}. \tag{3.3.1}$$

$$\int_{-W}^{W} e^{-\theta[(a-x)^2+(b-x)^2]} dx = \sqrt{\frac{\pi}{8\theta}} \left\{ \begin{array}{l} erf\left[\sqrt{2\theta}\left(W + \frac{a+b}{2}\right)\right] \\ +erf\left[\sqrt{2\theta}\left(W - \frac{a+b}{2}\right)\right] \end{array} \right\} e^{-\frac{\theta(a-b)^2}{2}}. \qquad (3.3.2)$$

$$\int_{x_{2;k}-\Delta}^{x_{2;k}+\Delta} e^{-\theta[(a-x)^2+(b-x)^2]} dx = \sqrt{\frac{\pi}{8\theta}} \left\{ \begin{array}{l} erf\left[\sqrt{2\theta}\left(\Delta - x_{2;k} + \frac{a+b}{2}\right)\right] \\ +erf\left[\sqrt{2\theta}\left(\Delta + x_{2;k} - \frac{a+b}{2}\right)\right] \end{array} \right\} e^{-\frac{\theta(a-b)^2}{2}}. \qquad (3.3.3)$$

### 3.4 A matrix identity

The trace of the product of $(n+1) \times (n+1)$, $(n \geq 1)$ matrices $A$ and $B$, with $B$ symmetric, equals the sum of their element-by-element products, i.e.,

$$tr(AB) = \sum_{i,j=1}^{n+1} A_{i,j} B_{i,j}. \qquad (3.4.1)$$

$$Demonstration: tr(AB) = tr\left(\sum_{j=1}^{n+1} A_{i,j} B_{j,k}\right) = \sum_{i=1}^{n+1}\sum_{j=1}^{n+1} A_{i,j} B_{j,i} = \sum_{i,j=1}^{n+1} A_{i,j} B_{i,j}.$$

## 4. Matrix $R$ for rectangular and circular-disk prediction domains

Our first order of business is to develop a means to compute the $IMSPE$ objective function of an $n$-point design over 2D, circular-disk prediction domains. We use Eq. 2.9 of the foundational paper [6] and the Appendix of [1] and set $\sigma_Z^2 = 1$, without loss of generality, so $IMSPE = 1 - tr(L^{-1}R)$, where $(n+1) \times (n+1)$ symmetric matrix $L$ is defined as $L \equiv \begin{pmatrix} 0 & 1 \\ . & V \end{pmatrix}$, and where $nxn$, symmetric matrix $V$ is the covariance matrix of the design $\mathcal{D} \equiv \{x_i; \ i = 1, \cdots, n\}$, having design components $(x_{i,j}; \ i = 1, \cdots, n; \ j = 1,2)$ [1]. Starting from closed-form elements of the $(n+1) \times (n+1)$ symmetric matrix $R$ for a rectangular prediction domain, we demonstrate a means to estimate $R$ for disks. Generalization to balls and hyper-balls is straightforward.

### 4.1 Matrix $R$ for rectangular prediction domains

Following [6,1] and referencing Fig. 4.1.1, below, we define matrix $R^{(r)}$ for a rectangular domain with center $(x_{1;k} = 0, x_{2;k})$, half-width $W$, half-height $\Delta$, and $v_i \equiv e^{-\theta_1(x_{i,1}-x_1)^2} e^{-\theta_2(x_{i,2}-x_2)^2}; \ i = 1, \cdots, n;$ as

$$R^{(r)} \equiv \frac{1}{4W\Delta} \int_{x_{2;k}-\Delta}^{x_{2;k}+\Delta} \int_{-W}^{W} \begin{pmatrix} 1 & v_1 & v_2 & \cdots & v_n \\ . & v_1^2 & v_1 v_2 & \cdots & v_1 v_n \\ . & . & v_2^2 & \cdots & v_2 v_n \\ . & . & . & \ddots & \vdots \\ . & . & . & . & v_n^2 \end{pmatrix} dx_1 dx_2, \qquad (4.1.1)$$

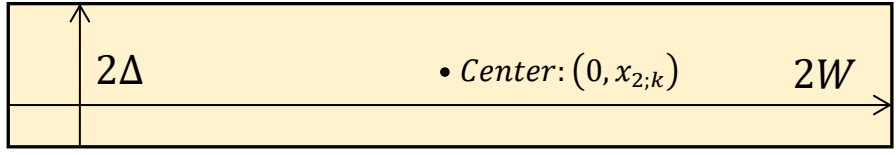

Fig. 4.1.1. The rectangular prediction domain has a center at $(0; x_{2:k})$, a horizontal width of $2W$, and a vertical height of $2\Delta$.

Using the integrals given in Eqs. 3.2.1, 3.3.1, 3.3.2, and 3.3.3, we obtain the upper-diagonal elements $(i \leq j)$ of $\boldsymbol{R}^{(r)}$, as follows:

$$R^{(r)}_{i \leq j} = \begin{cases} 1, & i, j = 0 \\ & (ULH\ corner) \\ \\ \dfrac{1}{4WD} \left( \sqrt{\dfrac{\pi}{4\theta_1}} \begin{Bmatrix} erf[\sqrt{\theta_1}(W + x_{j,1})] \\ +erf[\sqrt{\theta_1}(W - x_{j,1})] \end{Bmatrix} \cdot \sqrt{\dfrac{\pi}{4\theta_2}} \begin{Bmatrix} erf[\sqrt{\theta_2}(\Delta - x_{2;k} + x_{j,2})] \\ +erf[\sqrt{\theta_2}(\Delta + x_{2;k} - x_{j,2})] \end{Bmatrix} \right), & \begin{array}{l} i = 0\ and\ j = 1,\cdots,n \\ (top\ row) \end{array} \\ \\ \dfrac{1}{4WD} \left( \sqrt{\dfrac{\pi}{8\theta_1}} \begin{Bmatrix} erf\left[\sqrt{2\theta_1}\left(W + \dfrac{x_{i,1} + x_{j,1}}{2}\right)\right] \\ +erf\left[\sqrt{2\theta_1}\left(W - \dfrac{x_{i,1} + x_{j,1}}{2}\right)\right] \end{Bmatrix} \cdot \sqrt{\dfrac{\pi}{8\theta_2}} \begin{Bmatrix} erf\left[\sqrt{2\theta_2}\left(\Delta - x_{2;k} + \dfrac{x_{i,2} + x_{j,2}}{2}\right)\right] \\ +erf\left[\sqrt{2\theta_2}\left(\Delta + x_{2;k} - \dfrac{x_{i,2} + x_{j,2}}{2}\right)\right] \end{Bmatrix} \right) e^{-\left[\begin{array}{l}\theta_1(x_{i,1}-x_{j,1})^2 \\ +\theta_2(x_{i,2}-x_{j,2})^2\end{array}\right]/2}, & (otherwise), \end{cases}$$

(4.1.2)

which, agrees with the Appendix of [1] for the special case of a bi-unit square, viz., $W = \Delta = 1$, and $x_{2;k} = 0$. The elements for $i > j$ follow by symmetry.

## 4.2  Matrix $R$ for circular-disk prediction domains

For integration over the unit-radius circular disk, we seek,

$$\boldsymbol{R}^{(d)} \equiv \dfrac{1}{\pi} \int_{-1}^{1} \int_{-\sqrt{1-x_2^2}}^{\sqrt{1-x_2^2}} \begin{pmatrix} 1 & v_1 & v_2 & \cdots & v_n \\ \cdot & v_1^2 & v_1 v_2 & \cdots & v_1 v_n \\ \cdot & \cdot & v_2^2 & \cdots & v_2 v_n \\ \cdot & \cdot & \cdot & \ddots & \vdots \\ \cdot & \cdot & \cdot & \cdot & v_n^2 \end{pmatrix} dx_1\, dx_2. \quad (4.2.1)$$

These integrals are complicated because the inner integrals depend upon $x_2$, via the limits of integration, and because the outer integral is not of known, closed form. We use the following procedure to estimate the outer integral. We divide the disk into an even number $n_{int}$ of horizontally oriented rectangular sections; centered at $(0, x_{2;k})$, $k = 1, \cdots, n_{int}$, counting from the bottom of the disk, each with section height $2\Delta = 2n_{int}^{-1}$, as shown graphically in Fig. 4.2.1, below. It is evident that $x_{2;k}, k,$ and $\Delta$ are related by the following equation:

$$x_{2;k} = -1 - \Delta + 2k\Delta. \tag{4.2.2}$$

It remains to find an appropriate set of half-widths $W_k$, $k = 1, n_{int}$ for the rectangles approximating the sections. A simple choice is the horizontal distance to the ordinate from the intersection of the circle with the left-hand edge of the $k$'th section, at half-height, viz.,

$$W_k^{(simple)} \equiv \sqrt{1 - x_{2;k}^2}. \tag{4.2.3}$$

*Example* 4.2.1: With reference to Fig. 4.2.1, $n_{int} = 4, k = 3,$ and $x_{2;k} = x_{2;3} = -1/4$:

$$2W_3^{(simp.)} = 2\sqrt{1 - \left(\frac{1}{4}\right)^2} = \frac{\sqrt{15}}{2} = 1.936\cdots.$$

However, this method overestimates the area of each section, as represented graphically in Fig. 4.2.1 and as can be shown algebraically. A superior approach is to use the following, vertically averaged half-widths,

$$W_k \equiv \frac{1}{2\Delta}\int_{x_{2;i}-\Delta}^{x_{2;i}+\Delta} \sqrt{1 - x_{2;k}^2}\, dx_{2;k}. \text{ Using Eq. 3.1.1, with } a = 1,$$

$$W_k = \frac{n_{int}}{4}\left[x\sqrt{1-x^2} + \sin^{-1}(x)\right]\Big|_{x_{2;k}-n_{int}^{-1}}^{x_{2;k}+n_{int}^{-1}}$$

$$= \frac{n_{int}}{4}\begin{bmatrix}(x_{2;k} + n_{int}^{-1})\sqrt{1 - (x_{2;k} + n_{int}^{-1})^2} + \sin^{-1}(x_{2;k} + n_{int}^{-1}) \\ -(x_{2;k} - n_{int}^{-1})\sqrt{1 - (x_{2;k} - n_{int}^{-1})^2} - \sin^{-1}(x_{2;k} - n_{int}^{-1})\end{bmatrix}. \tag{4.2.4}$$

*Example* 4.2.2: Again with reference to Fig. 4.2.1:

$$n_{int} = 4, k = 3, x_{2;3} = -1/4, x_{2;k} = x_{2;3}, x_{2;3} + n_{int}^{-1} = 0, x_{2;3} - n_{int}^{-1} = -1/2:$$

$$2W_3 = 2\left[\frac{1}{2}\sqrt{1-\left(\frac{1}{2}\right)^2} - \sin^{-1}\left(-\frac{1}{2}\right)\right] = 2\left[\frac{\sqrt{3}}{4} + \frac{\pi}{6}\right] = \frac{3\sqrt{3} + 2\pi}{6} = 1.913\cdots.$$

We emphasize that $4W_k\Delta$ is the exact area of the $k$'th section of the disk.

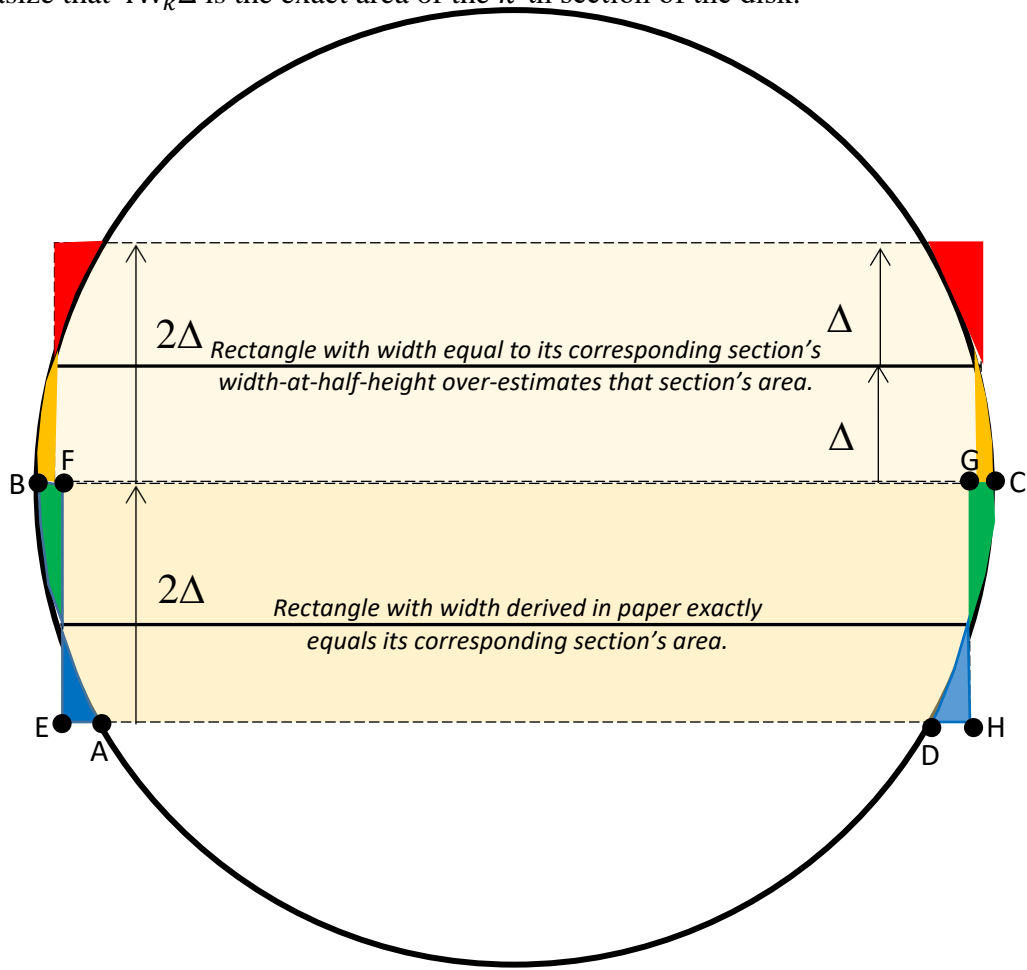

Fig. 4.2.1. The circular disk is the union of an even number $n_{int}$ ($n_{int} = 4$, in this figure) of non-overlapping sections, each having height $2\Delta$, a horizontal top and bottom, and right- and left-hand ends that are arcs of the bounding circle, e.g., Section ABCD, with obvious meaning. The required (integral) *IMSPE* over the disk is approximated by a sum over rectangles of height $2\Delta$, e.g., Rectangle EFGH. If the width of each such rectangle were taken as the width at half height of its corresponding section, each rectangle's area would overestimate the area of its corresponding section – shown as the red areas being larger than the orange ones. This paper uses a formula for the widths of the rectangles that gives the exact areas of their corresponding segments – shown as the green areas being equal to the blue ones.

Each upper-diagonal element of $\boldsymbol{R}^{(d)}$ is a sum of $n_{int}$ elements, via $R_{i,j}^{(d)} = \sum_{k=1}^{n_{int}} R_{i,j;k}^{(d)}$, where each of the $R_{i,j;k}^{(d)}$ is given by Eq. 4.1.2 for rectangles, but with W replaced by $W_k$ and overall normalization factor $1/(4W\Delta)$ replaced by $1/\pi$.

$$R_{i\leq j;k}^{(d)} = \begin{cases} 1, & i,j=0 \\ & (ULH\ corner) \\[2ex]
\dfrac{1}{\pi}\left(\sqrt{\dfrac{\pi}{4\theta_1}}\left\{\begin{array}{l}erf[\sqrt{\theta_1}(W_k + x_{j,1})]\\+erf[\sqrt{\theta_1}(W_k - x_{j,1})]\end{array}\right\}\cdot\sqrt{\dfrac{\pi}{4\theta_2}}\left\{\begin{array}{l}erf[\sqrt{\theta_2}(\Delta - x_{2;k} + x_{j,2})]\\+erf[\sqrt{\theta_2}(\Delta + x_{2;k} - x_{j,2})]\end{array}\right\}\right), & \begin{array}{l}i=0\ and\ j=1,\cdots,n\\(top\ row)\end{array} \\[4ex]
\dfrac{1}{\pi}\left(\sqrt{\dfrac{\pi}{8\theta_1}}\left\{\begin{array}{l}erf\left[\sqrt{2\theta_1}\left(W_k + \dfrac{x_{i,1}+x_{j,1}}{2}\right)\right]\\+erf\left[\sqrt{2\theta_1}\left(W_k - \dfrac{x_{i,1}+x_{j,1}}{2}\right)\right]\end{array}\right\}\cdot\sqrt{\dfrac{\pi}{8\theta_2}}\left\{\begin{array}{l}erf\left[\sqrt{2\theta_2}\left(\Delta - x_{2;k} + \dfrac{x_{i,2}+x_{j,2}}{2}\right)\right]\\+erf\left[\sqrt{2\theta_2}\left(\Delta + x_{2;k} - \dfrac{x_{i,2}+x_{j,2}}{2}\right)\right]\end{array}\right\}\right)e^{-\frac{\left[\theta_1(x_{i,1}-x_{j,1})^2+\theta_2(x_{i,2}-x_{j,2})^2\right]}{2}}, & (otherwise). \end{cases}$$

(4.2.5)

The elements for $i > j$ follow by symmetry.

## 5. Example: IMSPE-optimal, $n = 1$

We now show the $n = 1$, $IMSPE$-optimal design on the disk-shaped design domain, for the simple case $\theta_1 = \theta_2 = 1$, is simply the center point. We have $\boldsymbol{L} = \begin{pmatrix} 0 & 1 \\ \cdot & 1 \end{pmatrix}$, $\boldsymbol{R}^{(d)} = \begin{pmatrix} 1 & R_{0,1}^{(d)} \\ \cdot & R_{1,1}^{(d)} \end{pmatrix}$, and we seek to minimize $IMSPE = 1 - tr(\boldsymbol{L}^{-1}\boldsymbol{R})$, over design $\boldsymbol{x_1} = (x_{1,1}, x_{1,2})$. By inspection, $\boldsymbol{L}^{-1} = \begin{pmatrix} -1 & 1 \\ \cdot & 0 \end{pmatrix}$. Using Eq. 3.4.1,

$$IMSPE = 1 - tr(\boldsymbol{L}^{-1}\boldsymbol{R})$$

$$= 1 - \sum_{i,j=1}^{n+1}(L^{-1})_{i,j}R_{i,j}$$

$$= 2 - 2R_{0,1}^{(d)}. \tag{5.1}$$

We now compare the following three convergence methods for approximating the $IMSPE$, for this $n = 1$ problem, using $n_{int}$ evenly spaced, vertically stacked rectangles, as in Fig. 4.2.1, with centers at $x_{2;k}$, as given by Eq. 4.2.2; heights $2\Delta = 2n_{int}^{-1}$; and horizontal widths, as specified. For problems with $n > 1$, there are additional elements of $\boldsymbol{R}$ to be computed.

------

**Methods for approximating** $R_{0,1;k}^{(d)} = \dfrac{1}{\pi}\int_{-1}^{1}\int_{-\sqrt{1-x_2^2}}^{\sqrt{1-x_2^2}} e^{-\theta_1(x_{1,1}-x_1)^2 - \theta_2(x_{1,2}-x_{2;k})^2} dx_1 dx_2$

For all methods, the integrals are separable in $x_1$ and $x_2$, the normalization is $1/\pi$, and $R_{0,1}^{(d)}$ is approximated by $n_{int}$ rectangles with the following attributes:

    Centers: $(0, x_{2;k})$, with index $k$ (here and hereafter) running from $1, \cdots, n_{int}$;

    Widths (parallel to $x_1$ axis): $2W$ generically; and

    Heights (parallel to $x_2$ axis): $2\Delta = 2n_{int}^{-1}$.

*Convergence Method A*:

    Widths of rectangles: $2W_k^{(simple)}$, where $W_k^{(simple)}$ is given by Eq. 4.2.3;

    Inner integral: Given exactly by Eq. 3.2.1, with $W = W_k^{(simple)}$, $a = x_{1,1}$, and $x = x_1$;

    Outer integral approximation: Sum of $n_{int}$ terms with sizes $e^{-\theta_2(x_{1,2}-x_{2;k})^2} 2\Delta$; and

$$R_{0,1}^{(d)} = \lim_{n_{int}\to\infty} \frac{1}{\pi}\sum_{k=1}^{n_{int}} \sqrt{\frac{\pi}{4\theta_1}} \left\{\begin{array}{l} erf\left[\sqrt{\theta_1}\left(W_k^{(simple)} + x_{1,1}\right)\right] \\ + erf\left[\sqrt{\theta_1}\left(W_k^{(simple)} - x_{1,1}\right)\right]\end{array}\right\} e^{-\theta_2(x_{1,2}-x_{2;k})^2} 2\Delta.$$

*Convergence Method B*:

    Widths of rectangles: $2W_k$, where $W_k$ is given by Eq. 4.2.4;

    Inner integral: Given exactly by Eq. 3.2.1, with $W = W_k$, $a = x_{1,1}$, and $x = x_1$;

    Outer integral: Given exactly by Eq. 3.3.1, with $a = x_{1,2}$, and $x = x_2$.

$$R_{0,1}^{(d)} = \lim_{n_{int}\to\infty} \frac{1}{\pi}\sum_{k=1}^{n_{int}} \left(\begin{array}{l}\sqrt{\dfrac{\pi}{4\theta_1}}\left\{\begin{array}{l} erf[\sqrt{\theta_1}(\ W_k\ + x_{1,1})] \\ +erf[\sqrt{\theta_1}(\ W_k\ - x_{1,1})]\end{array}\right\} \\ \cdot \sqrt{\dfrac{\pi}{4\theta_2}}\left\{\begin{array}{l} erf[\sqrt{\theta_2}(\Delta - x_{2;k} + x_{1,2})] \\ +erf[\sqrt{\theta_2}(\Delta + x_{2;k} - x_{1,2})]\end{array}\right\}\end{array}\right).$$

*Convergence Method C*:

    Identical to Convergence Method A, but with $W_k^{(simple)}$ replaced by $W_k$:

$$R_{0,1}^{(d)} = \lim_{n_{int}\to\infty} \frac{1}{\pi}\sum_{k=1}^{n_{int}} \sqrt{\frac{\pi}{4\theta_1}} \left\{\begin{array}{l} erf[\sqrt{\theta_1}(W_k + x_{1,1})] \\ +erf[\sqrt{\theta_1}(W_k - x_{1,1})]\end{array}\right\} e^{-\theta_2(x_{1,2}-x_{2;k})^2} 2\Delta. \qquad (5.2)$$

------

For trial value $x_{1,1} = 0$, evaluation of Eq. 5.1, using Eq. 5.2 and Maple [7], for successively larger (even) values of $n_{int}$, followed by a convergence analysis to determine $IMSPE_0$, yields the log-log convergence plots in Fig. 5.1, below. Repeating for other trial values of $x_{1,1}$ yields a local-to-$x_{1,1} = 0$ parabola-shaped $IMSPE$ with minimum $IMSPE_0 = 0.735759\cdots$ at $x_{1,1} = 0$, i.e., a point at the origin. All results were checked by doubling the number of digits used in the Maple computations. No more than 160 digits of precision were required for these checks. The comparable 1D optimal objective function on the bi-unit interval is $IMSPE_0 = 0.506351\cdots$.

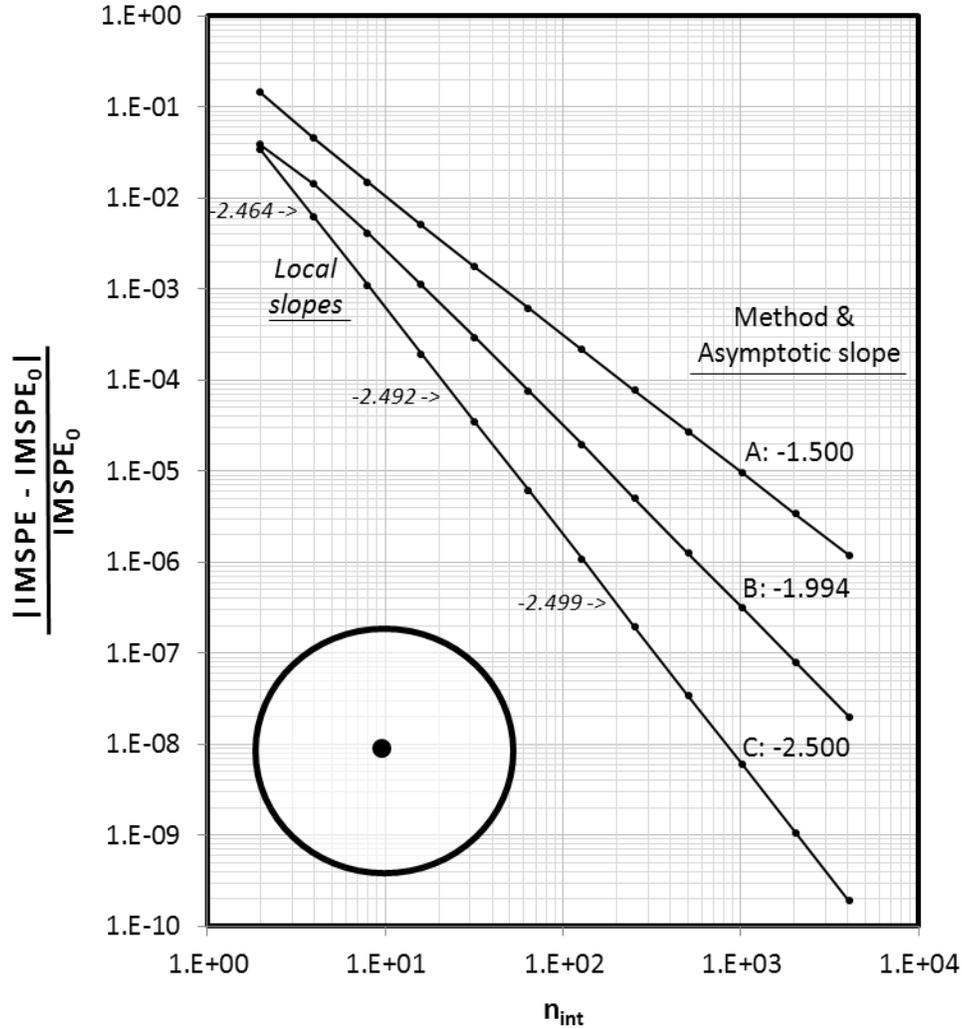

Fig. 5.1. (LLH-corner overlay): The singleton $IMSPE$-optimal design, on the unit-radius disk and assuming hyperparameters $\theta_1 = \theta_2 = 1$, is a centered point. (Log-log plot proper): For trial $x_{1,1} = 0$, which turns out to be the optimal design, small, black disks connected by straight-line segments show power-law convergence of the $IMSPE$ to $IMSPE_0 = 0.735758\cdots$, as the number of rectangles $n_{int}$ increases, under Convergence Methods A, B, and C described in the body of the paper. The (large-$n_{int}$) asymptotic slopes are shown, as are less-negative, local (to specific $n_{int}$'s) slopes, for Convergence Method C.

## 6. Example: IMSPE-optimal, n=2, on-axis centered design

We now find the $IMSPE$-optimal design, assuming $\theta_1 = \theta_2 = 1$, of two on-abscissa design points on the disk-shaped design domain, with centered midpoint, viz., $x_1 = (x_{1,1}; x_{1,2} = 0)$ and $x_2 = -x_1$, and taking $x_{1,1}$ as the single degree of freedom. In this case,

$$L = \begin{pmatrix} 0 & 1 & 1 \\ \cdot & 1 & A \\ \cdot & \cdot & 1 \end{pmatrix}, \text{where}$$

$$A = e^{-(x_{1,1}-x_{2,1})^2 - (x_{1,2}-x_{2,2})^2} = e^{-4x_{1,1}^2}, \text{and}$$

$$R^{(d)} = \begin{pmatrix} 1 & R_{0,1}^{(d)} & R_{0,2}^{(d)} \\ \cdot & R_{1,1}^{(d)} & R_{1,2}^{(d)} \\ \cdot & \cdot & R_{2,2}^{(d)} \end{pmatrix}.$$

By reflection symmetry about the $x_2$ axis, the Eq. 4.2.1 integrals of $v_1$ and $v_2$ are equal, as are the integrals of $v_1^2$ and $v_2^2$, so $R^{(d)}$ can be written in terms of just three of its elements, viz., $R_{0,1}^{(d)}$; $R_{1,1}^{(d)}$; and $R_{1,2}^{(d)}$. Then,

$$R^{(d)} = \begin{pmatrix} 1 & R_{0,1}^{(d)} & R_{0,1}^{(d)} \\ \cdot & R_{1,1}^{(d)} & R_{1,2}^{(d)} \\ \cdot & \cdot & R_{1,1}^{(d)} \end{pmatrix}.$$

We seek $IMSPE = 1 - tr(L^{-1}R)$.

By inspection, $L^{-1} = -\frac{1}{2}\begin{pmatrix} 1+A & -1 & -1 \\ -1 & -(1-A)^{-1} & (1-A)^{-1} \\ -1 & (1-A)^{-1} & -(1-A)^{-1} \end{pmatrix}$. Using Eq. 3.4.1,

$$IMSPE = 1 - \sum_{i,j=1}^{n} L_{i,j}^{-1} R_{i,j}$$

$$= 1 + \frac{1}{2}\left[(1+A) - 4R_{0,1}^{(d)} - 2R_{1,1}^{(d)}(1-A)^{-1} + 2R_{1,2}^{(d)}(1-A)^{-1}\right]$$

$$= \frac{3}{2} + \frac{e^{-4x_{1,1}^2}}{2} - 2R_{0,1}^{(d)} - \frac{R_{1,1}^{(d)}}{\left(1-e^{-4x_{1,1}^2}\right)} + \frac{R_{1,2}^{(d)}}{\left(1-e^{-4x_{1,1}^2}\right)}.$$

Using Eq. 4.2.5 gives

$$IMSPE = \frac{3}{2} + \frac{e^{-4x_{1,1}^2}}{2} - \frac{1}{8\sqrt{\theta_1\theta_2}} \sum_{k=1}^{n_{int}} \left[ \begin{array}{c} 4 \left( \begin{array}{c} \left\{ \begin{array}{c} erf[\sqrt{\theta_1}\ (W_k + x_{1,1})] \\ +erf[\sqrt{\theta_1}\ (W_k - x_{1,1})] \end{array} \right\} \\ \cdot \left\{ \begin{array}{c} erf[\sqrt{\theta_2}\ (\Delta - x_{2;k} + x_{1,2})] \\ +erf[\sqrt{\theta_2}\ (\Delta + x_{2;k} - x_{1,2})] \end{array} \right\} \end{array} \right) \\ + \frac{1}{(1-e^{-4x_{1,1}^2})} \left( \begin{array}{c} \left\{ \begin{array}{c} erf[\sqrt{2\theta_1}\ (W_k + x_{1,1})] \\ +erf[\sqrt{2\theta_1}\ (W_k - x_{1,1})] \end{array} \right\} \\ \cdot \left\{ \begin{array}{c} erf[\sqrt{2\theta_2}\ (\Delta - x_{2;k} + x_{1,1})] \\ +erf[\sqrt{2\theta_2}\ (\Delta + x_{2;k} - x_{1,1})] \end{array} \right\} \end{array} \right) \\ - \frac{1}{(1-e^{-4x_{1,1}^2})} \left( \begin{array}{c} \left\{ \begin{array}{c} erf[\sqrt{2\theta_1}\ (W_k)] \\ +erf[\sqrt{2\theta_1}\ (W_k)] \end{array} \right\} \\ \cdot \left\{ \begin{array}{c} erf[\sqrt{2\theta_2}\ (\Delta - x_{2;k})] \\ +erf[\sqrt{2\theta_2}\ (\Delta + x_{2;k})] \end{array} \right\} \end{array} \right) \end{array} \right],$$

(6.1)

where, just as in Sec. 5.1, $W_k$ is given by Eq. 4.2.4, $\Delta = n_{int}^{-1}$, and $x_{2;k}$ is given by Eq. 4.2.2.

An extension of Convergence Method C gives quadratic convergence to $IMSPE = 0.426149\cdots$, $x_{1,1} = -0.546820\cdots$, and $x_{2,1} = -x_{1,1}$, as shown in Fig. 6.1, below. The $IMSPE$ and optimality of the design are expected to be invariant under rigid-body rotation of the points about the origin, and this was confirmed. This design compares to the following $IMSPE$-optimal, two-point design, with hyperparameters $\theta_1 = \theta_2 = 1$, on the bi-unit design domain $[-1,1]^2$, found using methods described in Part I, Sec. 6 of this series of papers: $IMSPE^{(1D)} = 0.104338\cdots$, $x_{1,1}^{(1D)} = -0.547984\cdots$, and $x_{2,1}^{(1D)} = -x_{1,1}^{(1D)}$.

## 7. Example: IMSPE-optimal, $(n;\ \theta_1, \theta_2) = (4;\ 0.128, 0.00016)$ design

We now demonstrate a twin-point, $n = 4$, (putatively) $IMSPE$-optimal design, over coincident, circular-disk design and prediction domains. We start with the hyperparameters, viz., $\theta_1 = 0.128$ and $\theta_2 = 0.00016$, of one of the twin-point, $n = 4$, (putatively) $IMSPE$-optimal designs, over the coincident, square design and prediction domains, reported in [1] (and shown as Fig. 1.1 of the present paper). We assume there are two degrees of freedom, viz. $x_{1,1}$ and $x_{3,2}$, with the latter being the $x_2$-component of the locus of the third point. The remaining design coordinates are set by assuming the design is centered and has inversion symmetry: $x_2 = -x_1$ and $x_4 = -x_3$.

Analyses using Convergence Method C, carried out to give five significant figures for all $IMSPE$ values, demonstrate that all 4-in-line and all rectangular (including square) designs are sub-optimal, as shown graphically in Figs. 7.1 and 7.2, below. Further analyses show that the twin-point design with distal points $(\pm 0.69945\cdots, 0)$ and twin points $(0, \pm \delta)$, where $\delta = 0$ in the limit of infinite-precision optimization, has the lowest $IMSPE$, viz. $IMSPE_0 = 0.000029210\cdots$, of all rhomboid designs, thus demonstrating that non-round boundaries are not necessary for the appearance of twin-point (putatively) $IMSPE$-optimal designs. See Fig. 7.3 for a comparison of the inferred, putatively optimal, twin-point designs on circular-disk and square domains.

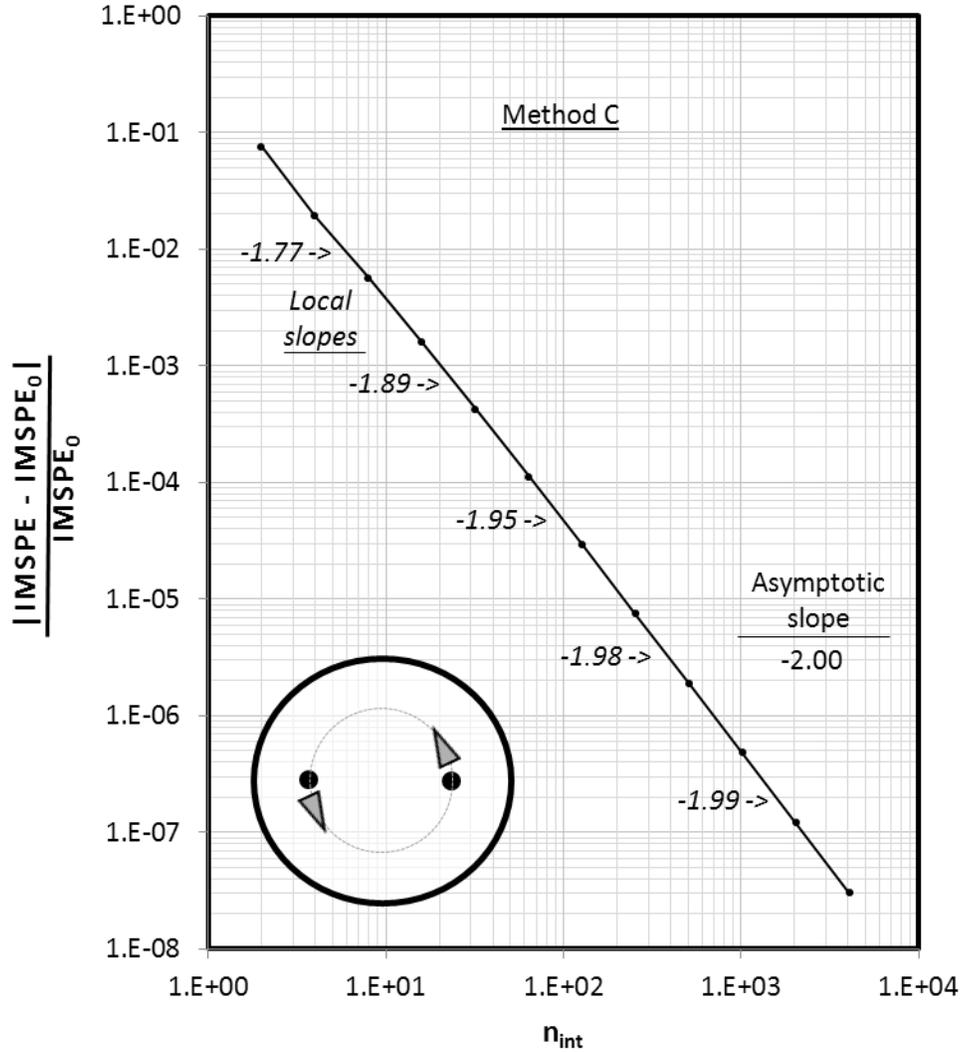

Fig. 6.1. (LLH-corner overlay): The two-point $IMSPE$-optimal design, on the unit-radius disk, assuming hyperparameters $\theta_1 = \theta_2 = 1$ and inversion symmetry about the origin, has points at radii $0.546820\cdots$. The design's continuous, counter-clockwise (or clockwise) symmetry is denoted by the grey arrows and dotted inner circle. (Log-log plot proper): Small, black disks connected by straight-line segments show convergence of the $IMSPE$ to $IMSPE_0 = 0.426149\cdots$, as the number of rectangles $n_{int}$ increases, under Convergence Method C, described in the body of the paper. The asymptotic slope for large $n_{int}$ is shown, as are smaller-negative, local (to $n_{int}$) slopes for moderate $n_{int}$.

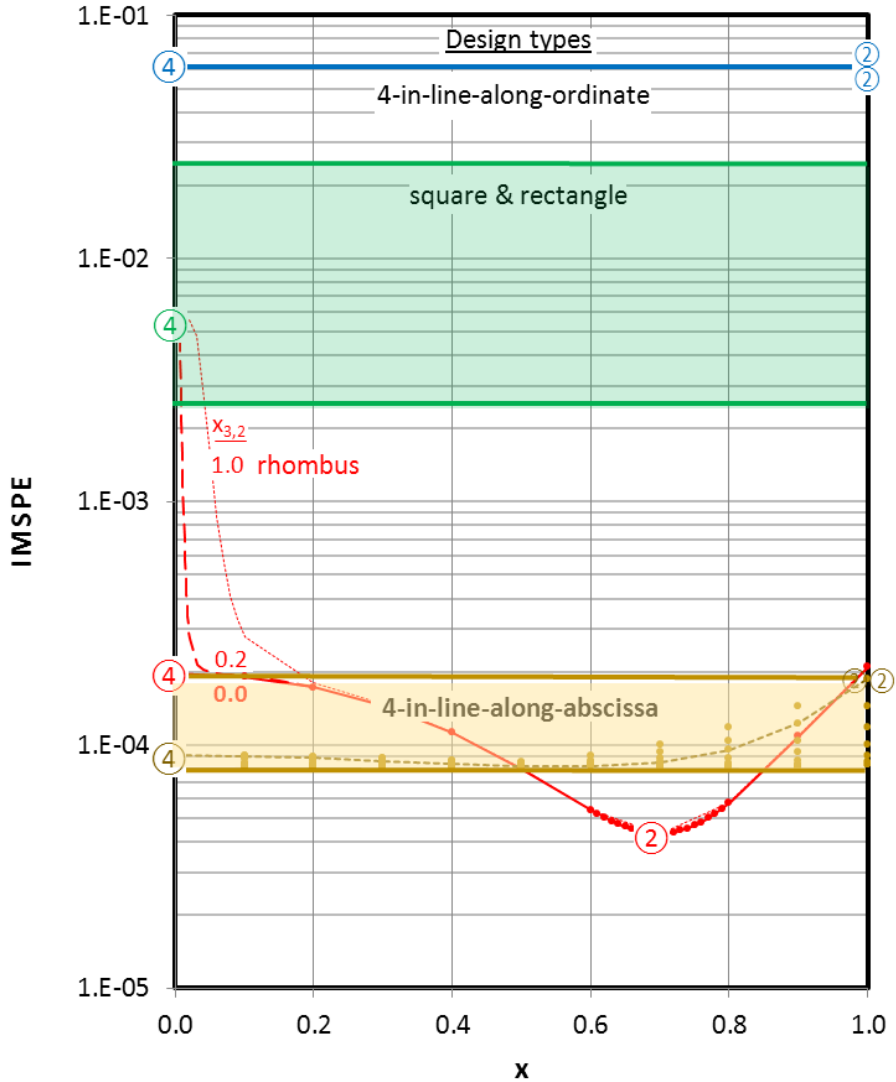

Fig. 7.1. The above plot shows bands of $IMSPE$ values for all centered, axis-oriented designs on the unit-circle disk. Designs are denoted in this caption using square brackets with meanings obvious from context. (Blue: 4-in-line designs along the ordinate, with abscissa label $x = x_{3,2}$): $IMSPE$'s lie in a band that includes the quadruplet-point design $[(0, \pm\delta); (0, \pm\varepsilon > \delta)]$, denoted by a ④, as well as a curve of "duet" twin-point designs, e.g., $[(0, \pm 0.7 \pm \delta)]$, ending at a blue duet ②; (Green: rectangular designs, with $x = x_{1,1} < x_{3,2}$): $IMSPE$'s lie in a band that includes the quadruplet-point design $[\pm\delta, \pm(\epsilon \geq \delta)]$ denoted by a ④; (Gold: 4-in-line designs along the abscissa, with $x = x_{1,1}$): $IMSPE$'s lie in a band that includes the quadruplet-point design $[(\pm\delta, 0); (\pm\varepsilon > \delta, 0)]$ denoted by a ④, as well as a dashed curve of duet twin-point designs, e.g., $[(\pm 0.7 \pm \delta, 0)]$, that ends at a gold double ②; (Red: rhomboid designs parameterized by $x_{3,2} = 0, 0.2$, or $1.0$, with $x = x_{1,1}$): $IMSPE$'s lie on solid, dashed, and dotted curves, respectively, closely overlap on the RHS and near the optimal, twin-point design at $[(\pm 0.69945 \cdots, 0); (0, \pm\delta)]$, denoted by a red ②, but split on the LHS into distinct risers, with the curve for $x_{3,2} = 0$ terminating at a red ④ and the other two curves peaking at $IMSPE \sim 0.0056$, before falling to local minima twin-point designs at $IMSPE \sim 0.50$, as shown in greater detail in Fig. 7.2.

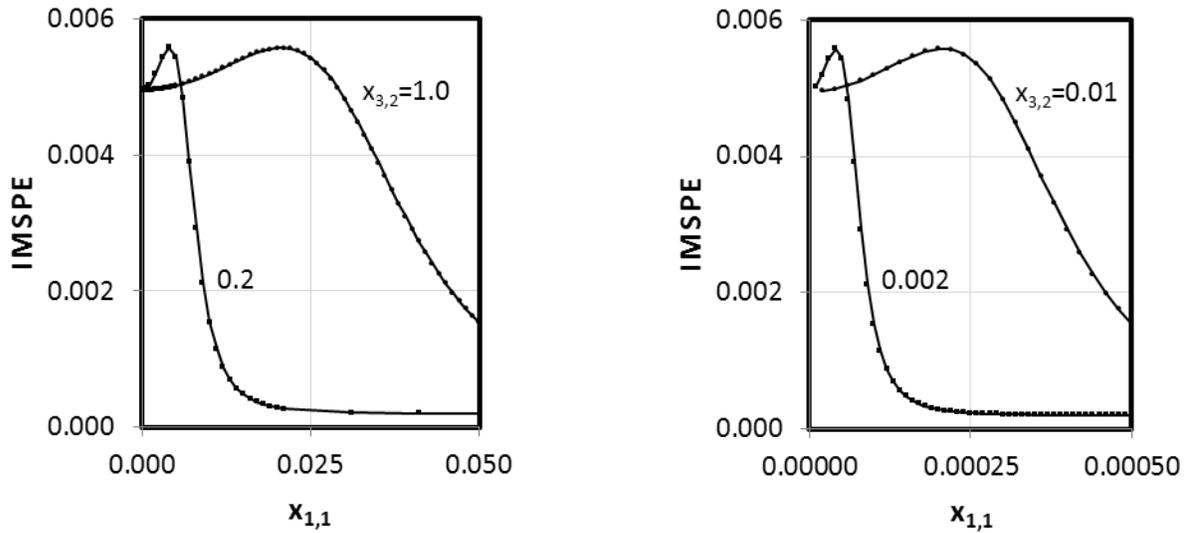

Fig. 7.2. (Left) Detail of the LHS of Fig. 7.1 shows how the quadruplet-point design at $IMSPE \sim 0.005$ is approached as $x_{1,1} \to 0^+$ for the two non-zero values of $x_{3,2}$ in that figure. (Right) Plotting the same graph with one-hundred-fold-diminished $x_{3,2}$'s and abscissa values gives a visually indistinguishable graph, demonstrating the invariances of the value and functional form of the $IMSPE$ multifunction, near the quadruplet-point design.

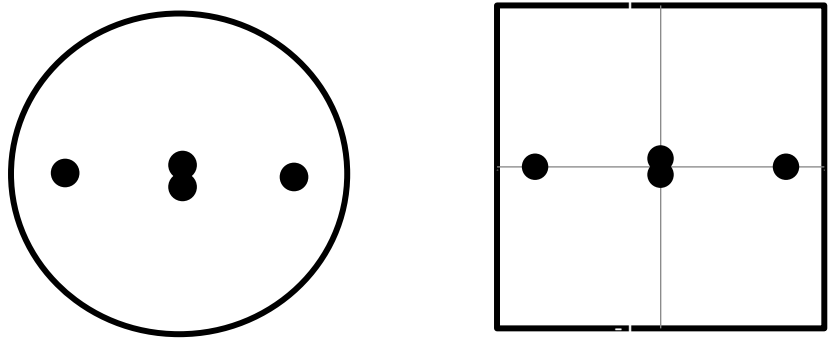

Fig. 7.3. (Left): The $n = 4$, twin-point, (putatively) $IMSPE$-optimal design for $\theta_1 = 0.128$ and $\theta_2 = 0.00016$ on the unit-radius circular disk has distal points $(\pm 0.69945 \cdots, 0)$; twin points $(0, \pm \delta)$, where $\delta = 0$ in the limit of infinite-precision design optimization; and $IMSPE = 0.000029210 \cdots$. (Right) The corresponding (putatively) $IMSPE$-optimal design on the bi-unit square [1] has distal points $(\pm 0..76711 \cdots, 0)$ and $IMSPE = 0.000066822 \cdots$.

The scale invariance of the $IMSPE$ evident in Fig. 7.2 is an attribute of Nu-class multifunctions that shall be explained in future parts of this series of papers.

## 8. Summary and concluding comments

We have demonstrated a method to evaluate the *IMSPE* on a coincident design and prediction domain on a circular disk. The method is extensible to more than two factors, to domains other than squares or disks, and to a variety of assumed covariance functions. We have discovered an $n = 4$, twin-point, (putatively) *IMSPE*-optimal design on such a domain.

## 9. Research reproducibility

We support the recommendations of ICERM's Workshop on Reproducibility in Computational and Experimental Mathematics Workshop [8]. All data and figure-generation files used in this research are available to responsible parties, upon request to selden_crary (at) yahoo (dot) com.

## 10. Revision history

V2: Eq. 4.2.2 was corrected. V3: Eq. 5.2 was removed. V4: Eqs. 4.1.2 and 4.2.5 were corrected. The last two equations of p. 11 and Eq. 6.1 were corrected. V5: "Generalized" was added to the title, and typos were corrected.

> *Anthropology demands open-mindedness with which one must look and listen, record in astonishment and wonder that which one would not have been able to guess.  --Margaret Mead*

## Acknowledgments

We thank Valerii Fedorov for his motivating Y2012 challenge regarding the possibility of clustered *IMSPE*-optimal designs for prediction domains with cuboidal vs. spheroidal boundaries. We also thank Prof. Ted Allen of Ohio St. Univ. for a comment about Version 2.